\documentclass[%
 reprint,
 amsmath,amssymb,
 aps,
]{revtex4-1}

\usepackage{graphicx}
\usepackage{dcolumn}
\usepackage{bm}

\begin{document}

\preprint{APS/123-QED}

\title{ Comments on the paper ``Terminal retrograde turn of rolling rings''}
\author{Alexey V. Borisov}
 \email{borisov@rcd.ru}
\author{Alexander A. Kilin}
 \email{aka@rcd.ru}
\affiliation{
 Udmurt State University, Izhevsk, Russia
}

\author{Yury L. Karavaev}
  \email{karavaev\_yury@istu.ru}
 \affiliation{
  Kalashnikov Izhevsk State Technical University, Izhevsk, Russia \\
  Udmurt State University, Izhevsk, Russia }

\date{\today}

\begin{abstract}
Mir Abbas Jalali et al. [Phys. Rev. E 92, 032913(2015)] explained the retrograde turn of rings by aerodynamic phenomena due to the presence of a central hole in the ring as opposed to a disk. The results of our experiments suggest that the drag torque is not the main reason for the retrograde turn of the rings, and the results of  theoretical research have shown that such a motion is possible for both the ring and the disk in the case of rolling without slipping.
\end{abstract}

\pacs{05.45.-a, 45.40.-f, 05.10.-a}
\maketitle

The authors of~\cite{Jalali} have discovered an intuitively unexpected fact, namely, that the rolling disk with a central hole makes a retrograde turn. In their opinion, the retragrade turn is caused by drag forces arising
from the fact that the disk has a central hole. But the model proposed in~\cite{Jalali}
provides no insight into the physical nature of the process of rolling of the ring.
Similar criticisms (made, e.g., in~\cite{Ruina, Engh, Pet}) concern also the paper by Moffatt~\cite{Mof}, who regards
the drag as the main mechanism of dissipation of energy of a homogeneous disk.
Nevertheless, the above-mentioned model has revived interest in the dynamics of a rolling disk and
generated a large amount of published research, the results of which cannot be ignored in the description of the retrograde turn of the ring.

References devoted to the study of the problem of the motion of a solid disk, which is known as the problem of Euler's disk,
can be found in~\cite{KO, Caps, Sae, Leine, BMKa, Ma_2016}.
The dynamics of Euler's disk is described using two models taking into account its contact with the surface: without
sliding~\cite{Kil} and with sliding~\cite{Engh, Prz}. The experiments reveal that sliding occurs at the initial
stages of motion and then changes to rolling motion. We recall that idealized sliding models are used to describe Hamiltonian systems,
while idealized rolling models are used for nonholonomic systems.
Within the framework of various friction models the dynamics of Euler's disk is sufficiently well understood, but some questions remain
regarding, for example, the loss of contact of the disk with the surface before it stops moving~\cite{BMKa}).
As the experiments show, the trajectory of a rolling homogeneous disk is different from the trajectory of a rolling ring in that
it contains no retrograde turn and is an undulating helix.
However, based on the results of theoretical studies, the authors of~\cite{Kil} have pointed out
that, depending on the level of energy and geometric and mass characteristics, the numerical nonholonomic model can
exhibit both retrograde and prograde rolling motion of the disk, but no transitions between the levels are possible
within the framework of this model.

It should be noted that the retrograde turn discovered by the authors of~\cite{Jalali} is not a new phenomenon
and is not typical only of the ring. It was observed earlier by A.C.Or in a numerical model developed to simulate
the dynamics of a tippe top~\cite{Or} and by R.Cross in his experiments with four different spinning tops~\cite{Cross}.
But the aerodynamics of the spinning tops considered in these papers differs significantly from that of a ring.

The refutation of the assertion that the air drag is the main cause of the retrograde turn of the ring during rolling is confirmed by simple experiments conducted by us. For the first experiment (Supplemental Material ~\cite{supvid1}) we have fabricated a special ring whose central hole has the form of a cylinder and the upper and lower bases of the ring have different radii (i.e., the wall of the ring in the diametral section has the form of a rectangular trapezoid). When the disk rolls on the larger base, the  trajectory of its motion resembles that of the disk and does not exhibit retrograde motions, and when the disk moves on the smaller base, no matter what the initial conditions, the ring executes a retrograde turn. The drag torque of the ring started on the
larger base differs insignificantly from the drag torque of the ring started on the smaller base, but the motion pattern
changes considerably. The results of this experiment allow the conclusion that the presence of retrograde turn of the ring depends strongly on the ratio between the radius of inertia of the ring and the radius of the base on which
the rolling motion occurs.

As a continuation of this experiment, we applied adhesive cellophane to one of the ring surfaces (Supplemental Material ~\cite{supvid2}).
This did not change considerably the geometric and mass characteristics of the ring, but the aerodynamics
approached most closely the characteristics of the disk. The results remained analogous to the previous experiment:
the ring made a retrograde turn only when it rolled on the smaller base. Similar experiments were carried out with
rings having other geometric and mass characteristics.
The independence of the trajectory of the rings from the presence of cellophane allows the conclusion
about a weak influence of air drag on the dynamics of the ring's rolling motion.

The third experiment was conducted in a vacuum chamber used for casting. The pressure in the chamber was reduced to the value $10^2 Pa$ (0,0145 psi). The experiments were carried out with a usual wedding gold ring,
which was thrown onto a horizontal aluminum plate by means of actuators installed in the vacuum chamber.
The phenomenon of retrograde turn of the ring in vacuum persisted, which is demonstrated by video file ~\cite{supvid3} (video file ~\cite{supvid4} shows the rolling of the same ring at atmospheric pressure). These experiments clearly show that the aerodynamic forces are not the main reason for the retrograde turn of the ring.

In this paper we present one of possible explanations for the phenomenon of the retrograde turn of a rolling disk with a central hole
using models taking into account rolling friction and constructed for Euler's disk.

To describe the dynamics of the retrograde motion, it is convenient to use a modified nonholonomic model
of a rolling disk~\cite{Kil} that takes into account rolling friction.
This model also describes the dynamics of the ring, which differs from the case of a disk only by different values of the mass and geometric parameters.
The practice of modifying idealized models provides a better qualitative explanation for some phenomena in a number of
problems, for example, in the tippe top problem~\cite{Cross} or in the rattleback problem~\cite{Tak}.
In spite of a wide range of experimental and theoretical data, modified models are a simple tool for a qualitative explanation
of phenomena arising during the motion of bodies.

In the nonholonomic model there exist both prograde and retrograde trajectories of the disk's motion~\cite{Kil}. The incorporation of dissipation into the nonholonomic model allows a transition from the prograde rolling to the retrograde turn. In this case, dissipation can be described by
different models. The most complete research results on the mechanisms of energy dissipation of a rolling disk,
including experimental results, are presented in~\cite{Leine, Ma_2014}. The authors of these papers believe that
after the short sliding stage the dynamics of Euler's
disk is mainly influenced by rolling friction, which in~\cite{Ma_2014, Ma_2016} is simulated by viscous contact with
quadratic dependence on the velocity of motion.
This is verified by a good agreement of numerical simulation results for the nutation angle and the precession velocity with
the experimental research results. However, the authors of~\cite{Leine, Ma_2014} do not pay proper attention to the trajectories
of the disk under the conditions of the friction models under study nor to the dependence of the shape of the trajectories on
the geometric and mass characteristics of the disks.
We note that the model of viscous contact was first introduced by Contensou in~\cite{Cont}.

\begin{figure}[!ht]
\centering
\center{\includegraphics[width=0.8\linewidth]{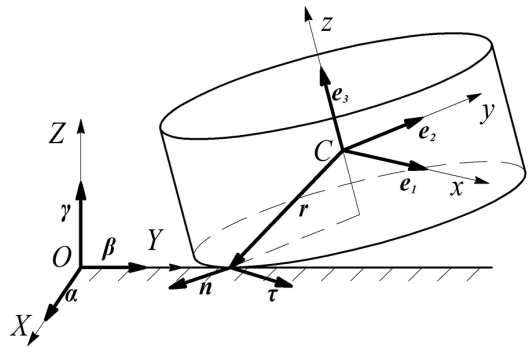}}
\caption{Geometry of a rolling disk}
\label{fig_disk}
\end{figure}

To describe the motion of the disk, we consider two coordinate systems (see Fig.~\ref{fig_disk}): a fixed system, $OXYZ$,
with the unit vectors $\boldsymbol{\alpha}, \boldsymbol{\beta}, \boldsymbol{\gamma}$, and a moving system, ${Cxyz}$, rigidly attached
to the center of mass of the disk, with the unit vectors $\boldsymbol{e_1}, \boldsymbol{e_2}, \boldsymbol{e_3}$.
We specify the position of the system by the coordinates of the center of mass of the disk in the fixed coordinate system
$\boldsymbol{r_c}=(x, y, z)$ and by the matrix defining the orientation of the disk in the space $\mathbf{Q}=\left( \boldsymbol{\alpha}, \boldsymbol{\beta},  \boldsymbol{\gamma} \right)$.
Here and in the sequel (unless otherwise stated) all vectors are referred to the moving coordinate system ${Cxyz}$.

The equations of motion of the disk can be represented as
\begin{equation}
\begin{gathered}
\boldsymbol{\mathrm{I}}\dot{\boldsymbol{\omega}} = \boldsymbol{\mathrm{I}}\boldsymbol{\omega}\times\boldsymbol{\omega} + \boldsymbol{r}\times\boldsymbol{N} + \boldsymbol{\mathrm{M_f}} \\
m \dot{\boldsymbol{v}} = m \boldsymbol{v}\times\boldsymbol{\omega} - mg\boldsymbol{\gamma} + \boldsymbol{N}.
\label{eq_d}
\end{gathered}
\end{equation}
where $ \boldsymbol{\mathrm{M_f}}$ and $\; \boldsymbol{N}$ are, respectively, the moment of rolling friction and the reaction force
of the plane, $m$ is the mass of the disk, and $\boldsymbol{r}$ is the radius vector of the point of contact in the moving
coordinate system
\begin{equation}
\boldsymbol{r} = \left(- \frac{R \gamma_2}{\sqrt{1-{\gamma_3}^2}}, - \frac{R \gamma_1}{\sqrt{1-{\gamma_3}^2}}, -h\right),
\label{eq_r}
\end{equation}
where $R$ is the radius of the base of the disk on which it rolls, and $h$ is the distance from this base to the center of mass.

The absence of slipping at the point of contact is ensured by the nonholonomic constraint
\begin{equation}
\boldsymbol{f} = \boldsymbol{v} + \boldsymbol{\omega}\times\boldsymbol{r} = 0.
\label{eq_f}
\end{equation}

From the common solution of equations (\ref{eq_d}) and the time derivative of the constraint (\ref{eq_f}) one can find the
reaction force $\boldsymbol{N}$. Substituting the resulting expression into the first equation of (\ref{eq_d}) and eliminating
the velocities using the constraint equation (\ref{eq_f}), we obtain

\begin{equation}
\boldsymbol{\mathrm{I}}\dot{\boldsymbol{\omega}} = \boldsymbol{\mathrm{I}}\boldsymbol{\omega}\times\boldsymbol{\omega} + m\boldsymbol{r}\times (\boldsymbol{r}\times \dot{\boldsymbol{\omega}} + \dot{\boldsymbol{r}}\times\boldsymbol{\omega} - (\boldsymbol{r}\times\boldsymbol{\omega})\times\boldsymbol{\omega} + g\boldsymbol{\gamma}) + \boldsymbol{\mathrm{M_f}}.
\label{eq_eq}
\end{equation}

For physical considerations it is convenient to decompose the moment of rolling friction into three components
corresponding to friction torques in different directions (see also~\cite{Ma_2014}): the friction torque preventing the disk from
spinning relative to the vertical $\boldsymbol{\gamma}$
with the friction coefficient $\mu_{\gamma}$, the friction torque preventing the disk from rotating about the tangent vector
$\boldsymbol{\tau}$ to the edge of the disk at the point of contact with the surface with the friction coefficient
$\mu_{\tau}$, and the friction torque preventing the disk from rolling along the edge of its lower base with the friction
coefficient $\mu_{n}$. In the general case, when the three friction coefficients are different, the friction torque can be
represented as

\begin{equation}
\boldsymbol{\mathrm{M_f}} = - \boldsymbol{\mathrm{\widehat{\mu}}}\boldsymbol{\omega} , \;\;\;\; \boldsymbol{\mathrm{\widehat{\mu}}} = \mu_{\gamma}\boldsymbol{\gamma}\otimes\boldsymbol{\gamma} + \mu_{\tau}\boldsymbol{\tau}\otimes\boldsymbol{\tau} + \mu_{n}\boldsymbol{n}\otimes\boldsymbol{n},
\label{eq_M}
\end{equation}
where the unit tangent vector $\boldsymbol{\tau}$ and the unit normal vector $\boldsymbol{n}$ at the point of contact are
\begin{equation*}
\boldsymbol{\tau} = \frac{\boldsymbol{\gamma}\times\boldsymbol{e_3}}{\sqrt{1-{\gamma_3}^2}}, \;\; \boldsymbol{n} = \frac{\boldsymbol{\gamma}\times(\boldsymbol{e_3}\times\boldsymbol{\gamma})}{\sqrt{1-{\gamma_3}^2}},
\end{equation*}
and the tensor product of the vectors $\boldsymbol{a}, \boldsymbol{b}$ is defined as follows:
$$
\boldsymbol{a}\otimes \boldsymbol{b} = \|a_i b_j \|.
$$

In the general case, the friction coefficients $\mu_{\gamma}$, $\mu_{\tau}$, and $\mu_{n}$ can depend on the phase variables and
the reaction force of the plane $\boldsymbol{N}$. In this paper, we restrict ourselves to the case of viscous rolling friction
linear in angular velocities. In this case, all coefficients $\mu_{\gamma}$, $\mu_{\tau}$, and $\mu_{n}$ are constant.
To simplify the model, we also assume that all three coefficients
are equal to $\mu_{\gamma} = \mu_{\tau} = \mu_{n} = \mu$ and that the moment of rolling friction takes the form
$\boldsymbol{\mathrm{M_f}} = - \mu\boldsymbol{\omega}$. It turns out that even under this fairly rough assumption the results
of numerical simulation are in a good qualitative agreement with the experimental results.

Adding to (\ref{eq_eq}) the kinematic Euler equations and quadratures for the center of mass of the disk, we obtain
a closed system of equations governing the dynamics of the disk
\begin{eqnarray}
&&\boldsymbol{\mathrm{I}}\dot{\boldsymbol{\omega}} + \boldsymbol{\omega}\times\boldsymbol{\mathrm{I}}\boldsymbol{\omega} + \nonumber\\ m(\boldsymbol{r}\times\dot{\boldsymbol{\omega}} + &&\dot{\boldsymbol{r}}\times\boldsymbol{\omega} - (\boldsymbol{r}\times\boldsymbol{\omega})\times\boldsymbol{\omega} + g\boldsymbol{\gamma})\times\boldsymbol{r} = - \mu\boldsymbol{\omega}, \nonumber \\
\dot{\boldsymbol{\alpha}} = &&\boldsymbol{\alpha}\times\boldsymbol{\omega}, \;\; \dot{\boldsymbol{\beta}} = \boldsymbol{\beta}\times\boldsymbol{\omega}, \;\;\dot{\boldsymbol{\gamma}} = \boldsymbol{\gamma}\times\boldsymbol{\omega}, \nonumber\\
&&\dot{x} = (\boldsymbol{\alpha}, \boldsymbol{r}\times\boldsymbol{\omega}), \;\; \dot{y} = (\boldsymbol{\beta}, \boldsymbol{r}\times\boldsymbol{\omega}).
\label{eq_sys}
\end{eqnarray}
Below we present results of the numerical simulation of equations (\ref{eq_sys}) and their comparison with
the experimental results.

\begin{figure}[!ht]
\centering
\begin{minipage}[!ht]{0.5\linewidth}
\center{\includegraphics[width=0.95\linewidth]{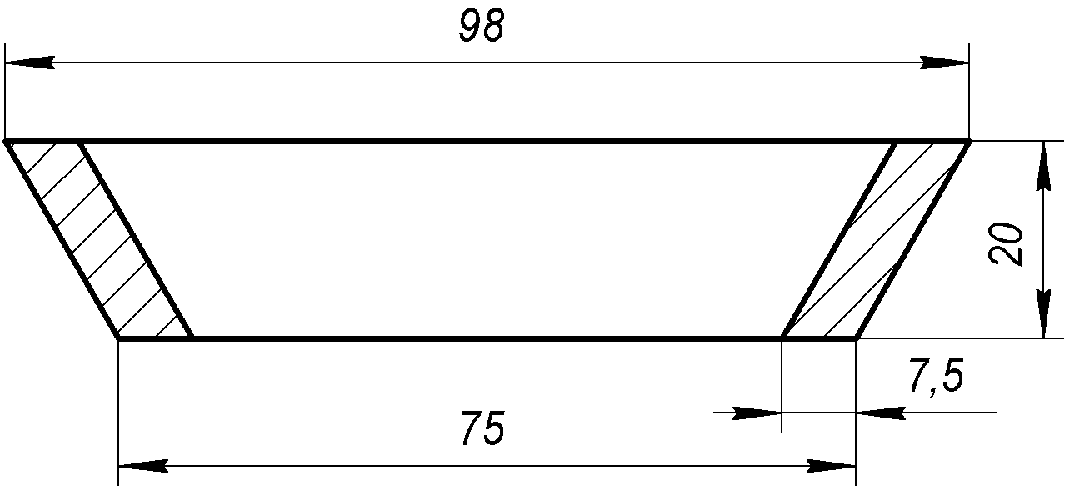} \\ a)}
\end{minipage}
\begin{minipage}[!ht]{0.45\linewidth}
\center{\includegraphics[width=0.9\linewidth]{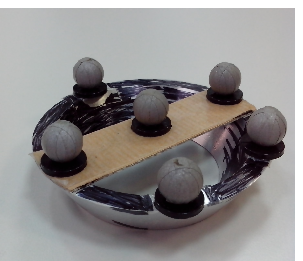}\\ b)}
\end{minipage}
\caption{(a) Sketch of the experimental model. (b) Photograph of the experimental model with markers of the motion capture system.} \label{fig_mark}
\end{figure}

To carry out experimental research, we have developed an experimental model, a ring of radius variable in height,
whose diametral section is shown in Fig. \ref{fig_mark}a. The parameters of the ring have the following values:
radii of the bases: $R_1 =0.0375$ m, $R_2 =0.049$ m, mass: $m = 0.1034$ kg, axial moments of inertia:
$I_x = I_y = 0.08647\cdot10^{-3}$ kg $\cdot$m$^2$, $I_z = 0.1661\cdot10^{-3}$ kg $\cdot$m$^2$, height of the center of mass:
$h =0.0105$ m. The experimental trajectory of the center of mass was restored by means of a motion capture system with frequency
200 Hz using the markers on the ring (see Fig. \ref{fig_mark}b).

\begin{figure*}[!ht]
\centering
\begin{minipage}[!ht]{0.45\linewidth}
\center{\includegraphics[width=0.6\linewidth]{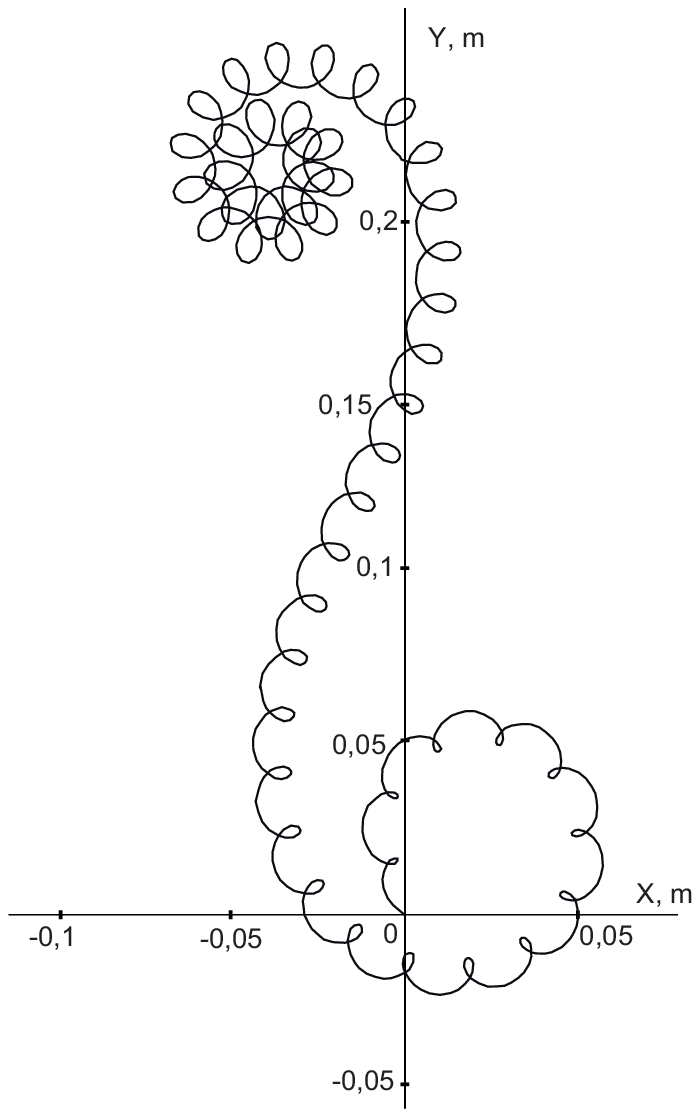} \\ a)}
\end{minipage}
\begin{minipage}[!ht]{0.45\linewidth}
\center{\includegraphics[width=0.7\linewidth]{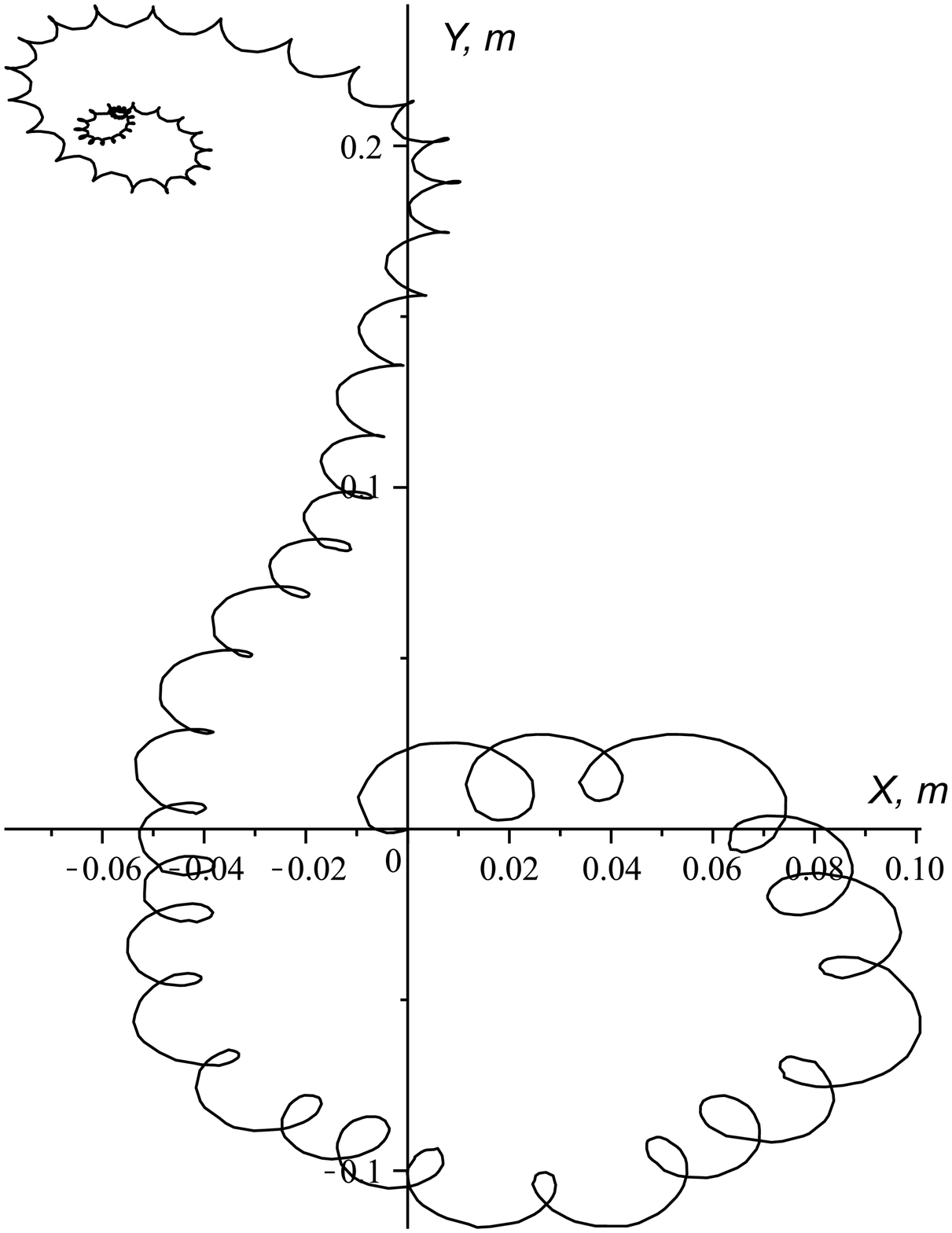}\\ b)}
\end{minipage}
\caption{a) Trajectory of a rolling ring with rolling friction (numerical simulation), b) Trajectory of a rolling
ring, restored by means of the motion capture system.} \label{fig_1}
\end{figure*}

When the experimental model rolled on the larger base, no retrograde turns were observed under different initial conditions.
When it rolled on the smaller base, a retrograde turn occurred under initial conditions corresponding
to an energy level higher than a critical one.
An example of an experimental trajectory of the rolling ring on the smaller base is given in Fig. \ref{fig_1}b.

Using the motion capture system, we have obtained the initial conditions $\omega_1(0) = -2.5119$, $\omega_2(0) = 7.6737$, $\omega_3(0) = 29.2722$,
$\gamma_3 = 0.6964$ for this trajectory, which were used for the numerical simulation. By virtue of the axial symmetry of
the disk and freedom in the choice of a fixed coordinate system, the other initial
conditions can be given arbitrarily. The trajectory of the rolling ring which was obtained by numerical simulation of equations
(\ref{eq_sys}) under the above initial conditions with the friction coefficient $\mu = 5.17\times10^{-6}$ is presented in Fig. \ref{fig_1}a.

As can be seen from Fig. \ref{fig_1}, the trajectory is qualitatively in good agreement with the experiment. The size and the number of loops on the trajectory (small loops of the cycloids) depends strongly
on the coefficient of friction, and their match with the experiment can be achieved by introducing nonuniform rolling friction into the model.
The results of numerical simulation show that the solution of the system is sensitive to the initial conditions and the values
of the friction coefficients. However, qualitatively the trajectory is the same as that obtained
in the experiment in a wide range of geometric and mass parameters of the disk and initial conditions.

In this paper we have restricted ourselves to considering one model of viscous rolling friction, since the results of
research on dissipation mechanisms presented in~\cite{Leine, Ma_2014} have shown a key role of this model and
good agreement with the experiment. Nevertheless, we note that the retrograde turn described above was experimentally observed
also in the case of rolling with slipping over some portions of the trajectory. When the model
of the disk rolling with slipping was applied for the whole duration of motion, no retrograde turn of the ring was observed.
Analysis of the model with slipping for particular segments of the trajectory is a more complicated problem, since
the transition from sliding to rolling requires taking dry friction into account, which can lead to paradoxical
phenomena~\cite{MIv, IvM} and requires separate research. Therefore, the results of quantitative research are unpredictable
and will hardly be obtained in the near future. Theoretically, it seems impossible to construct a trajectory that
is quantitatively in agreement with the experiment, since there are microroughnesses and the characteristics of the materials
are inhomogeneous, which is important in problems with friction.

To conclude, we summarize the results obtained in our work.

1. Air drag is not a reason for the retrograde turn of the ring during its rolling.

2. The phenomenon of the retrograde turn of the ring can be explained qualitatively within the framework of the model of
a rolling ring with viscous rolling friction.

3. Since the transitions between rolling friction and sliding friction which arise on some segments of the trajectory are
accidental and paradoxical, it is difficult to construct a trajectory of the ring that would be quantitatively the same as
the experimental trajectory.

\begin{acknowledgments}
The computer simulation was carried out using the software package "Computer Dynamics: Chaos" (http://lab-en.ics.org.ru/lab/page/kompyuternaya-dinamika/). The theoretical research carried out by A.V. Borisov and supported by the RSF grant no. 15-12-20035.  The work of A.A. Kilin was supported by the RFBR grant no.15-08-09261-a. The work of Yu.L. Karavaev was supported by the RFBR grant no. 15-38-20879 mol\_a\_ved.
\end{acknowledgments}

\bibliography{bibfile}

\end{document}